\documentclass[]{aa}
\usepackage{graphics}
\usepackage{graphicx}
\usepackage{natbib}
\usepackage{txfonts}
\usepackage{balance}
\usepackage[a4paper]{hyperref}
\begin{document}
\idline{1}{0}
\doi{0}
\def\year{2005}

\def\teff{$T\rm_{eff}$ }
\def\kms {$\mathrm{km\, s^{-1}}$ }
\def\vsini {$\mathrm{v\,sin\,i$}}

\title{ Sulphur abundances in Terzan 7 
\thanks{Based on data  obtained in ESO programme 65.L-0481}
}


   \author{E. Caffau \inst{1}
\and
P. Bonifacio \inst{2}
\and
R. Faraggiana \inst{3}
\and
L. Sbordone \inst{4,5}
}
\offprints{E. Caffau}
\institute{
Liceo L. e S.P.P. S. Pietro al Natisone, annesso al
Convitto Nazionale ``Paolo Diacono'',
Piazzale Chiarottini 8, Cividale del Friuli (Udine), Italy 
   \email{elcaffau@libero.it}
\and
Istituto Nazionale di Astrofisica - Osservatorio Astronomico di
Trieste,
    Via Tiepolo 11, I-34131
             Trieste, Italy\\
   \email {bonifaci@ts.astro.it}
\and
Dipartimento di Astronomia, Universit\`a degli Studi di Trieste
   \email{faraggiana@ts.astro.it}
\and
ESO European Southern Observatory -
              Alonso de Cordova 3107 Vitacura, Santiago, Chile
             \and
             Universit\'a di Roma 2 ''Tor Vergata'' -
             via della Ricerca Scientifica, Rome, Italy
\\
   \email{sbordone@mporzio.astro.it}
}

\authorrunning{Caffau et al.}
\titlerunning{Sulphur in Terzan 7}
\date{Received ...; Accepted ...}

\abstract{

We present here the first measurements of sulphur abundances
in extragalactic stars. We make use of high resolution spectra,
obtained with UVES at the ESO 8.2 m Kueyen telescope, of three
giants of the Globular Cluster Terzan 7, which belongs
to the Sagittarius dwarf galaxy. We measure the sulphur abundances
using the lines of \ion{S}{i} multiplet 1.
The S/Fe ratios for all three stars are nearly solar, thus
considerably lower than what is found in Galactic stars
of comparable iron content ([Fe/H]$\sim -0.50$).
This finding is in keeping with the abundances
of other $\alpha$-chain elements in this cluster and in 
Sagittarius and other dSphs in general.
These low $\alpha$-chain elements to iron ratios 
suggest that Sagittarius and its Globular Clusters 
have experienced a low or bursting star-formation rate.
Our sulphur abundances imply
$\langle \log (S/O)\rangle = -1.61$
which is comparable to what is 
found in many \ion{H}{ii} regions of
similar oxygen content, and is slightly  lower 
than the solar value (log (S/O)\sun $= -1.51$).
These are also the first measurements of sulphur abundances
in a Globular Cluster, thus a direct comparison of Terzan 7
and Galactic Globular Clusters is not possible yet. However
our analysis suggests that the lines of \ion{S}{i} multiplet 1 should
be measurable for other Globular Clusters at least down to
a metallicity $\sim -1.5$.

\keywords{Nucleosynthesis - Stars: abundances -
globular clusters: individual: Terzan 7  
- Galaxies: abundances -
Galaxies: dwarf -  Galaxies:individual Sgr dSph }}

\maketitle

\section{Introduction}

\begin{figure*}[tb]
   \centering
\includegraphics[width=\hsize,clip=true]{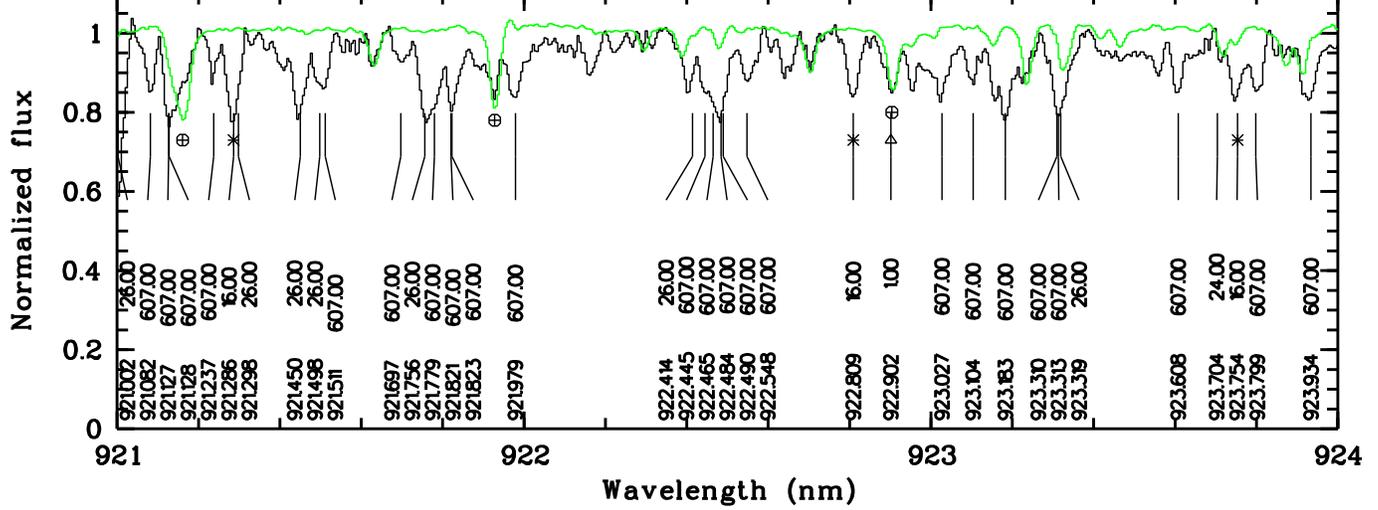}
\caption{
The spectrum of star 
 \# 1282 (black line) and the spectrum of the B star
HD  68761 ($v\sin i = 350$ \kms) (grey line),
which has been used to remove the telluric lines.
The three most prominent telluric features have been marked
with a crossed circle. The three sulfur lines (marked by asterisks), 
as well as the other stellar lines, with central residual intensity less than
0.8
have been
identified. Also the position of Paschen $\zeta$ has been identified
(and marked by a triangle), 
although its predicted  central residual intensity is 0.9665; note
that, due to the radial velocity of Terzan 7 this coincides with the 
telluric line at 923.48 nm.  
}
\label{tel}
\end{figure*}

\begin{figure}[tb]
   \centering
\includegraphics[width=8cm,clip=true]{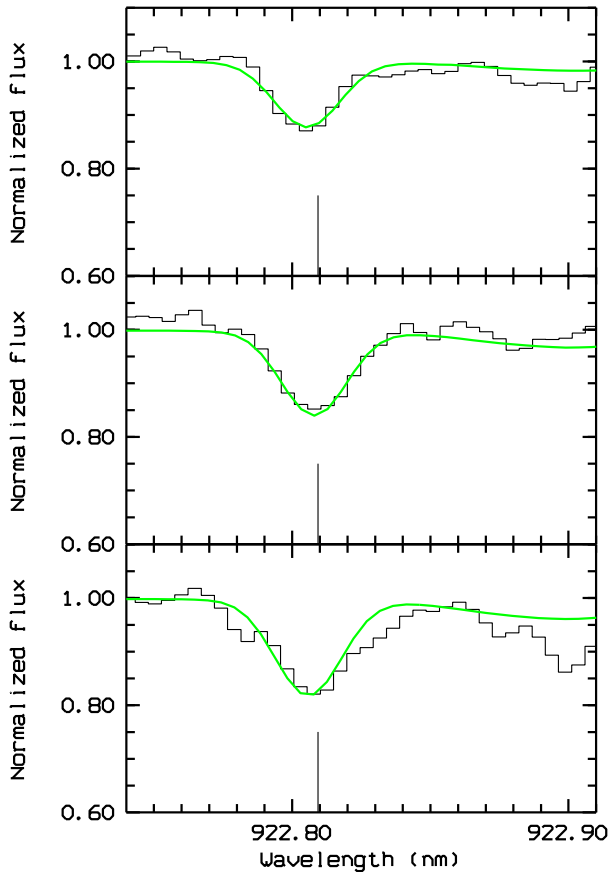}
\caption{Fit of the 922.8 nm line of the three stars, from top to bottom
are stars \# 1665, \# 1282  and \# 1708.
This line is not contaminated by telluric lines in
any of the stars.}
\label{fit}
\end{figure}

The $\alpha$-chain elements (i.e. all the even elements from oxygen 
to titanium) may be produced in stars
during carbon-burning, oxygen-burning
and neon-burning phases, both in central burning and convective shell
burning, as well as in explosive burning phases \citep{limongi}.
It is only the massive stars, which end their lives as 
Type II SNe, which undergo these phases and have a means
to eject their nucleosynthesis products in the interstellar medium.
Thus the $\alpha$-chain elements are a good tracer of
the nucleosynthesis of the short lived massive stars.
On the other hand Fe and other iron-peak elements 
are produced both in Type II SNe and in 
Type Ia SNe, which may explode over  longer time-scales
\citep{tinsley,chiaki}. 
For this reason the ratios $\alpha$/Fe (where $\alpha$
is any of the $\alpha$-chain elements) are sensitive
diagnostics which may give us information on the
time scales for the evolution of a galaxy and on
the star formation rate,
although their interpretation is not always
straightforward (see \citealt{chiaki} and references
therein).
Among the $\alpha$-chain elements
sulphur is not often studied 
in stars because there are few suitable lines, 
at variance with the neighbouring elements Si and Ca
which are more easily measured.
Sulphur is instead more easily measured in the 
interstellar medium, both in the warm ISM, through
absorption lines 
\citep[][and references therein]{savage_sembach} and in 
\ion{H}{ii} regions through
emission lines \citep{garnett,tp89}.
This makes  sulphur an element which is readily 
measured in external galaxies in which
one of its gaseous phases is measurable, i.e. Damped
Ly $\alpha$ galaxies (DLAs,\citealt{centurion}) and
Blue Compact Galaxies (BCGs,\citealt{garnett,IT99}).
With respect to other easily accessible
$\alpha$ elements, such as Si or Mg,
sulphur has the advantage that it is not 
depleted onto dust grains \citep{savage_sembach};
thus its abundance in the gas phase equals
the total abundance.

\begin{table}
\caption{Atmospheric parameters and S abundances.}
\label{result}
\begin{center}
{\scriptsize
\begin{tabular}{rrrlcccc}
\hline
\hline
\\
 \multispan2{\hfill Star\hfill}   &\teff & log g  &$\xi$  & [Fe/H] & [S/H]  & [S/Fe]\\
      & &        & cgs    &\kms   & dex    &  dex   &  dex   \\
\hline
\hline
 1665 & S16 &3945  & 0.8    & 1.55   & --0.51 & --0.59 & --0.08\\
 1282 & S34 &4203  & 1.3    & 1.60   & --0.54 & --0.59 & --0.05\\
 1708 & S35 &4231  & 1.2    & 1.70   & --0.56 & --0.62 & --0.01\\
\hline                                                                                      
\end{tabular}
}
\end{center}
\end{table}

This situation makes it highly desirable to have
a direct comparison with sulphur abundances measured
in stars, either in the Milky Way or in gas-poor galaxies,
such as dwarf spheroidals, for which stellar measurements
are the main, or only source of abundances.
Although it could be argued that other $\alpha$-chain
elements  such as Ca
or Si could be used as proxies, there are some theoretical
predictions that not all $\alpha$-chain elements should 
vary in lockstep \citep[][and references therein]{lanfranchi},
as well as some observational hints \citep{venn}.
Clearly, accurate observations of several $\alpha$-chain
elements are needed to decide if this is the case or not.
For these reasons the additional effort to measure
sulphur in stellar spectra is justified. 

Terzan 7 \citep{terzan} 
is a 
Globular Cluster 
associated with the Sgr (dSph) system.
Its low stellar concentration allowed to obtain accurate photometry into
the central region; from these data the young age and the metallicity
have been estimated \citep{buonanno95}. However it appeared soon that
the metallicity determined photometrically (lower than [Fe/H]= --0.74,
\citealt{buonanno95}) is
in clear disagreement with that obtained spectroscopically from 
\ion{Ca}{ii} 
triplet lines 
([Fe/H]$=-0.36\pm 0.11$ \citealt{dacosta}). 
The metallicity  derived from high
resolution spectra of giant stars 
is [Fe/H]=--0.61 according to \citet{T04}
and [Fe/H]=--0.59 according to 
\citet{S05},
both obtained from
spectra observed with UVES. 

Both \citet{T04} and \citet{S05} found that the abundance 
of $\alpha$-chain elements, notably Ca, Si and Mg, implies
$\alpha$  to iron ratios which are {\em lower} than the
ratios observed in Galactic stars of comparable
metallicity. It is thus quite interesting to investigate
whether sulphur behaves like the other $\alpha$-chain elements.
We  stress  that these are the  first
measurements of S in a Globular Cluster.
The usually studied 
$\alpha$ elements are: O, Si, Ca and Ti.

\section{Data analysis and results}

\begin{figure}
   \centering
\includegraphics[width=\hsize,clip=true]{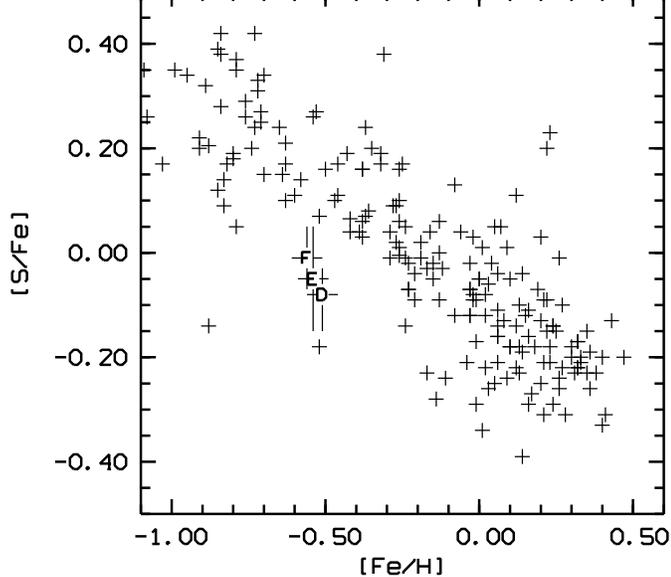}
\caption{[S/Fe] versus [Fe/H] for Galactic stars of the compilation of
\citet{caffau} (crosses) and Terzan 7;  each star is
 denoted by a letter: \# 1665 is D, \# 1282 is E and
\# 1708 is F.}
\label{fes}
\end{figure}

Our sample consists of three giant stars of Terzan 7
observed with UVES at the 8.2 m Kueyen ESO telescope
which have already been analyzed
by \citet{T04} and \citet{S05}.
UVES has been used with dichroic \# 2, for the present
work we used only the data of the upper CCD in the
red arm. The resolution is R$\sim 43000$.
We determine the sulphur abundance using line profile fitting 
in the region from 921 nm to 924 nm (see Fig. \ref{tel}), 
which
covers the three lines of the \ion{S}{i} multiplet 1, 
the signal to noise ratio in this region is $\sim 50$ 
for all three stars.
We use a $\chi^2$ minimization code
as we did for sulphur determination for Galatic stars
\citep[see][]{caffau}.
Telluric lines were subtracted making
use of the spectrum of a fast rotator, suitably
scaled. In all three stars one sulphur line
(922.8 nm) was not blended with telluric lines and this 
allowed to check the success of 
the telluric subtraction procedure.
The 921.2 nm line has very near a telluric line, while
the 923.8 nm line was contaminated.

With  respect to the Galatic stars which
we analyzed \citep{caffau}, these Terzan 7
stars are cooler and of lower gravity, 
so that some weak CN molecular lines are present. 
Of the three  sulphur lines only the 923.8 nm is blended
with a CN feature, however for all three the presence
of these CN lines affects the position of the continuum.
Therefore  at first we fitted the CN features in the
region between 921.2 nm and 922.8 nm, in order to fix the CN abundance,
then we fitted S lines.
Fits to the 922.8 nm line in the three stars are shown
in Fig. \ref{fit}. 
We do not report here the adopted abundances of C and N because
we are not sure of the quality of the oscillator strengths
we are using. The abundances of C and N in these stars
will be the object of future work, making use also of the blue
spectra available. 
For the stars we adopted the 
atmospheric parameters and metallicities
derived by \citet{S05}, 
these are reported in Table \ref{result} together 
with the derived 
sulphur abundances, the star numbers refer
to the \citet{buonanno95} catalogue, the names used by \citet{T04}
are also provided.
The difference in S abundance derived  from the three different lines
is $\le 0.06$ dex in all cases. 
The model atmospheres were the same used by
\citet{S05},  the 
effective temperatures were derived from the $B-V$ colour
and have an uncertainity of the order of 100 K. 
The model atmosphere for star \# 1665 was computed 
with ATLAS 12  and custom abundances, while
for the other two stars ATLAS 9 with  solar-scaled 
Opacity Distribution Functions
for $\xi = 1$ \kms was used.
The synthetic spectra
were computed using the {\tt SYNTHE} suite
\citep{kurucz} in its Linux version
\citep{S04}. The oscillator strengths
of the \ion{S}{i} lines were taken from
\citet{wiese}.

To estimate the errors in the sulphur abundances
we resorted to a Monte Carlo simulation.
Since all the stars have similar atmospheric
parameters and signal to noise ratios,
we performed the simulation only for
star  \# 1282, and take these error
estimates as representative also for
the other stars.
A Monte Carlo set is obtained by injecting
noise into a synthetic spectrum so that S/N=50,
all sets comprised  10000 events.
The input synthetic spectrum had 
\teff = 4203 K, log g=1.30, $\xi = 1.60$ \kms [Fe/H]=--0.54,
[S/Fe]=--0.05. 
The simulated spectrum is fitted as though it
were an observed spectrum and the standard deviation 
from the mean fitted abundance is taken as 
error estimate.
The synthetic spectra  used  in the  fitting have 
either equal
\teff, log g and $\xi$  as the input
spectrum, in order to estimate the  random error
due to the noise in the data,
or 
different parameters (\teff, log g, $\xi$,
[Fe/H]) to estimate the joint effect of systematic errors
in the parameters and noise in the data. 

From these simulations we derive a random error
of 0.04 dex. A change of metallicity of 0.2 dex
results in a change of about 0.10 dex in the mean
S abundance (+0.10 for an increase of 0.20 in [Fe/H];
--0.06 for a decrease  of 0.20 in [Fe/H]). 
A change in log g  of $\pm0.5$ dex results in
a change of $\pm 0.18$ dex in sulphur abundance.
A change in \teff of $\pm 100$ K results in a
change in  sulphur 
abundance of $\mp 0.18 $ dex
(note the change in sign).

\begin{figure}
   \centering
\includegraphics[width=\hsize,clip=true]{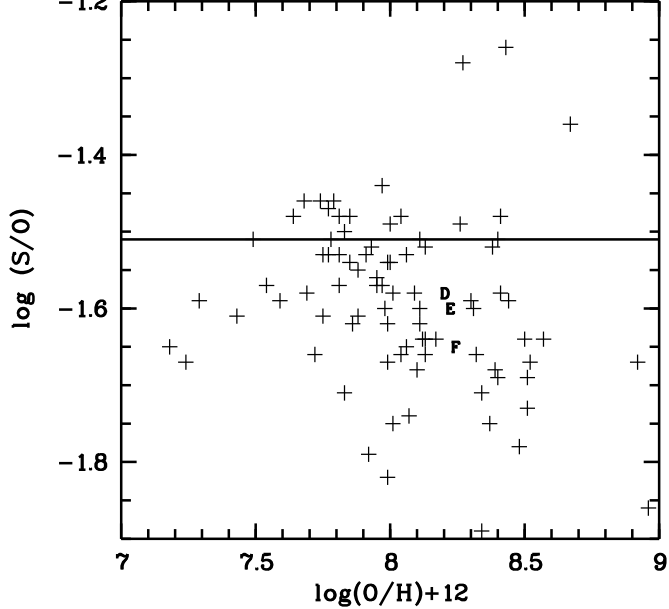}
\caption{Comparison of the S/O ratio in the stars of
Terzan 7 (each star is designated by a letter
as in Fig. \ref{fes}) and in \ion{H}{ii} regions,
both Galactic and extragalactic (crosses), the
latter data are from \citet{garnett}
and \citet{IT99}. The solid
line represents the solar value:
A(S)\sun = 7.21 \citet{lodders}
and A(O)\sun=8.72 \citet[][from MARCS 1D models]{asplund} }
\label{s_o}
\end{figure}

\section{Discussion}

With respect to Galactic stars 
of comparable metallicity, the three Terzan 7 stars which we
analyzed are clearly deficient in S. In Fig. \ref{fes} are shown  
the [S/Fe] ratios versus [Fe/H] for the stars of the compilation of 
\citet{caffau}, and each of our Terzan 7 stars is
identified with a  
letter. Virtually all Galactic stars, of metallicity 
comparable to Terzan 7, have higher [S/Fe] ratios. 
We note a few Galactic stars with low [S/Fe] ratios,
which could be  accreted by the Galaxy,
however their Galactic orbits do not appear to be
distinctive \citep{caffau}.  
Our main conclusion 
is that sulphur appears to track the other
$\alpha$ elements. Moreover, since for the other
$\alpha$ elements, Terzan 7 seems to be undistinguishable
from the Sgr field stars of similar metallicity, one may
expect the same to hold for sulphur: a prediction which may
be verified with observations of Sgr field stars.
This conclusion relies on the accuracy of our effective 
temperatures, if these were systematically higher by\balance
100 - 150 K the [S/Fe] ratios in Terzan 7 would become
similar to those of Galactic stars. The good agreement
between our colour temperatures and the excitation temperatures
of \citet{T04} suggests that such a systematic shift
is unlikely. 
The [S/Fe] ratios found by us seem to strengthen
the similarity of Sagittarius with DLA galaxies,
which had been already asserted on the basis
of other $\alpha$ elements \citep{B04}.

Our results may also be directly compared with those
of \citet{garnett} and \citet{IT99} who concluded that
in BCGs the S/O ratio is constant, independent of the 
metallicity. This has been taken as evidence that in
such galaxies the initial mass function (IMF) does not vary
with time and that the same massive stars are responsible both
for the production of oxygen and sulphur. 
The oxygen abundances in our stars have been 
measured by \citet{luca_tesi} and the S/O
ratios are shown in Fig. \ref{s_o} compared
to the values in \ion{H}{ii} regions, both Galactic
and extragalactic, from \citet{garnett} and \citet{IT99}.
The solar value is shown as a horizontal line.
The stars in Terzan 7 occupy a position which is populated
by many \ion{H}{ii}  regions,
slightly below the solar
S/O ratio.
From  Fig. \ref{s_o} we note a rather large
scatter, which might be due to observational errors,
however it is intriguing that most galaxies (and Galactic 
\ion{H}{ii} regions)
show S/O ratios which are {\em below} the solar
value.
Note that in the previous work on S abundances in 
\ion{H}{ii} regions
this fact was not apparent because the value A(O)\sun~=~8.93 for
the solar oxygen abundance, from \citet{anders}, was adopted.
Instead we have adopted the determination of 
\citet[][using MARCS 1D models]{asplund}.
 The quality of the data does not allow to
claim that IMF variations actually exist. 
However new and more accurate observations of S/O
ratios, both in stars and \ion{H}{ii} regions should 
be able to   address this point.
The success of our measurement of S in Terzan 7
suggests that the lines of multiplet 1 should
be measurable, at least down to [S/H]$\sim -1.5$,
for stars in Globular Clusters and Local Group galaxies.

\bibliographystyle{aa}

\begin{thebibliography}{}


\bibitem[Anders \& Grevesse(1989)]{anders} Anders, E. \& 
Grevesse, N.\ 1989, Geochim. Cosmochim. Acta 53, 197



\bibitem[Asplund et al.(2004)]{asplund} Asplund, M., Grevesse, 
N., Sauval, A.~J., Allende Prieto, C., \& Kiselman, D.\ 2004, \aap, 417, 
751 

\bibitem[{Bonifacio et al. (2004)}]{B04}
Bonifacio, P., Sbordone, L., Marconi, G., Pasquini, L., \& Hill,~V., 2004, A\&A
414, 503


\bibitem[{Buonanno et al. (1995)}]{buonanno95}
Buonanno R., Corsi, C. E., Pulone, L.,
Fusi Pecci, F., Richer,~H. B., \& Fahlman, G. C. 
1995, AJ 109, 663

\bibitem[Caffau et al.  (2005)]{caffau}
Caffau E., Bonifacio P., Faraggiana R., Fran\c cois P., 
Gratton~R.G., \& Barbieri M.
\ 2005, \aap, submitted

\bibitem[Centuri{\' o}n et al.(2000)]{centurion} Centuri{\' o}n, 
M., Bonifacio, P., Molaro, P., \& Vladilo, G.\ 2000, \apj, 536, 540 



\bibitem[Da Costa \& Armandroff(1995)]{dacosta} Da Costa, 
G.~S., \& Armandroff, T.~E.\ 1995, \aj, 109, 2533 


\bibitem[Garnett(1989)]{garnett} Garnett, D.~R.\ 1989, \apj, 
345, 282 


\bibitem[Izotov \& Thuan(1999)]{IT99} Izotov, Y.~I., \& 
Thuan, T.~X.\ 1999, \apj, 511, 639 


\bibitem[Kobayashi et al.(1998)]{chiaki} Kobayashi, C., 
Tsujimoto, T., Nomoto, K., Hachisu, I., \& Kato,~M.\ 1998, \apjl, 503, L155 

\bibitem[{Kurucz(1993)}]{kurucz}
Kurucz R. L., 1993, CDROM 13, 18,
\href{http://kurucz.harvard.edu/}{http://kurucz.harvard.edu/}


\bibitem[Lanfranchi \& Matteucci(2003)]{lanfranchi} Lanfranchi, 
G.A., \& Matteucci, F.\ 2003, \mnras, 345, 71 





\bibitem[Limongi \& Chieffi(2003)]{limongi} Limongi, M.~\& 
Chieffi, A.\ 2003, MSAIS, 3, 58 

\bibitem[Lodders(2003)]{lodders} Lodders, K.\ 2003, \apj, 591, 
1220 



\bibitem[Savage \& Sembach(1996)]{savage_sembach} Savage, B.~D., \& 
Sembach, K.~R.\ 1996, \araa, 34, 279 

\bibitem[Sbordone (2005) ]{luca_tesi}
Sbordone, L., 2005, PhD Thesis, Universit\'a di Tor Vergata


\bibitem[{Sbordone et al. (2004) }]{S04}
Sbordone, L., Bonifacio, P., Castelli, F., \& Kurucz, R. L., 2004, MSAIS, 5, 93

\bibitem[Sbordone et al. (2005)]{S05} Sbordone L., Bonifacio P.,
Marconi G., Buonanno R., \& Zaggia~S.\  2005, \aap, in press

\bibitem[{Tautvai\v sien\' e et al. (2004)}]{T04}
Tautvai\v sien\' e, G., Wallerstein G., Geisler D., Gonzales, G., \& Charbonnel, C., 2004, AJ, 127, 373

\bibitem[Terzan(1968)]{terzan} Terzan, A.\ 1968,
C.R.~Acad.~Sci.~Ser.~B1, 267, 1245



\bibitem[Tinsley(1979)]{tinsley} Tinsley, B.~M.\ 1979, \apj, 
229, 1046 

\bibitem[Torres-Peimbert et al.(1989)]{tp89} 
Torres-Peimbert, S., Peimbert, M., \& Fierro, J.\ 1989, \apj, 345, 186 

\bibitem[Venn et al.(2004)]{venn} Venn, K.~A.,
Irwin, M.,
Shetrone, M.~D., Tout, C.~A., Hill, V., \& Tolstoy, E.
\ 2004, \aj, 128, 1177



\bibitem[Wiese et al.(1969)]{wiese} Wiese, W.~L., Smith, 
M.~W., \& Miles, B.~M.\ 1969, NSRDS-NBS, Washington, D.C.: US Department of 
Commerce, National Bureau of  Standards  



\end{thebibliography}

\end{document}